# Characterization techniques of Fe-doped CuO thin films deposited by the Spray Pyrolysis method

Fatima Zahra CHAFI[1*], Lahoucine BAHMAD[2], Najem HASSANAIN[1], Boubker FARES[1], Larbi LAANAB[3] and Ahmed MZERD[1]

1. Laboratory of Physics and Materials (LPM) - Mohammed V University, Faculty of Science- BP. 1014- Av. Ibn Batouta Rabat, Morocco
2. Laboratory of Magnetism and Physics of High Energy (LMPHE-PPR-13) - Mohammed V University, Faculty of Science – BP. 1014- Av. Ibn Batouta Rabat, Morocco
3. Laboratory of Conception and Systems (LCS) – Mohammed V University, Faculty of Science – BP. 1014- Av. Ibn Batouta Rabat, Morocco

*Corresponding author. chafifatimazahra@gmail.com

## ABSTRACT

The Fe-doped CuO thin films were deposited onto glass substrates by Spray pyrolysis technique. The structural, micro-structural, optical and electrical properties of the synthesized samples were investigated in details. The X-Ray diffraction (XRD), Raman and Fourier Transform Infrared (FTIR) spectroscopy, confirmed that the studied samples exhibit single phase monoclinic structure of CuO. The UV-VIS spectrophotometer mentioned that the transmittance increases to 80 % when increasing the Fe concentration. Furthermore, the band gap energy of the obtained CuO was 1.29 eV. This value was slightly increased by the Fe substitution. In addition, the electrical properties of the films such as the conductivity, the mobility, the resistivity and the carrier concentration have been studied. The Hall Effect measurements confirmed the p-type conductivity of the studied films.

## 1. Introduction

Because of their sizes, morphology and structure, the metal oxides nanoparticles have been intensively studied in the last decade [1–3]. Copper oxide nanoparticles have attracted extensive attention recently, because of their unique properties. Indeed the CuO has been studied as a p-type semiconductor with a band gap value 1.2 eV, because of the natural abundance, low cost production processing, nontoxic nature, and its reasonably good electrical and optical properties [4-13].

Because of its interesting properties, the Copper oxide material is an important candidate for a variety of practical applications. These properties are included in the electronic and the optoelectronic devices, such as micro-electromechanical systems, spintronics, solar cells, gas sensors [14] and photo thermal applications. Also one find these practical applications in the high temperature super-conducting materials, diodes, lithium ion electrochemical cell [15], field emission devices [16], nano-fluid [17], giant magneto resistance materials [18, 19] and nano-devices for catalysis [20].

To synthesize the CuO compound, many different chemical and physical methods have been reported, such as Sol gel, chemical vapor deposition, electro-deposition, thermal oxidation, sputtering process and spray pyrolysis [21-24].

The present investigation deals with the study of Iron doping on Copper Oxide thin films by spray pyrolysis technique. The effect of iron doping on various samples of CuO thin films on the structural, Raman, optical and electrical properties have been studied and discussed. In section 2, we illustrate the experimental methods. Section 3 is devoted to results and discussions.

## 2. Experimental methods

Undoped and Fe-doped Copper Oxide thin films were prepared onto glass substrates at 350°C by spray pyrolysis technique. A homogeneous solution (200ml) was prepared by dissolving 0.1M of ($CuCl_2.2H_2O$) and ($FeCl_2.2H_2O$) in deionized water at room temperature. The solutions were stirred with a magnetic stirrer for 1hour to obtain a clear solution. The glass substrates were previously cleaned in Acetone and Methanol, rinsed with large amount of distilled water then dried to remove any residual moisture and placed over a hot plate. The temperature of the samples was monitored by a chromel-alumel thermocouple close to the substrates and was maintained at 350°C. The nozzle was located approximately at 40 cm above the substrate. The solution was sprayed into fine droplets using air as a carrier gas. In order to avoid any contamination, air was compressed from the atmosphere using filter to remove water and dust. Spray rate was held at 1.6 ml/min. Different solutions were prepared with different concentration amount of Fe (0, 2, 5, 10, 15 at %), the additives were controlled to obtain the following compositions; CuO (pure), $Cu_{0.98}Fe_{0.02}O$, $Cu_{0.95}Fe_{0.05}O$, $Cu_{0.90}Fe_{0.10}O$, $Cu_{0.85}Fe_{0.15}O$. These compositions are named: $Cu_{1-x}Fe_xO$ (x=0.00; 0.02; 0.05; 0.10; 0.15) respectively.

The films we studied were characterized by different techniques. An attempt has been carried out to examine the feasibility of thin films. The obtained X-Ray diffraction patterns were recorded by a system called D8 advance Bruker, using CuKα-radiation of wavelength (λ= 0.154056 nm). The scanning we performed for the 2θ angle in the range of 10°- 60°. The Raman method was investigated with FT-Ram 40-100 mW, vertex 70 within resolution 4 with 64 scans. Concerning the optical transmittance spectra measurements, we obtained them using the UV-visible spectrophotometer-Perkin Elmer lambda 900, using the wavelength range 400-1400 nm. Finally, the electrical properties were determined by the ECOPIA Hall Effect measurements using the Van Der Paw configuration at room temperature in air.

3. Results and discussion

3.1. Structural properties

The structural characterization of the samples was carried out by X-Ray diffraction (XRD) patterns in the range of 2θ between 10° and 60°. Figure 1 shows the typical diffraction peaks which indicate that the samples are well crystallized. Two highly intense diffraction peaks are observed at values 35.65° and 38.84° with a preferred orientation respectively along the (-111) and (111) axis and which can be indexed as monoclinic structured CuO. Other less intense peaks are obtained at 32.56°, 48.96°, 53.72° and 58.40° corresponding to (110), (-202), (020) and (202) respectively, when increasing the Fe content. The specific crystallographic planes confirmed the formation of CuO monoclinic (tenorite) phase that belongs to the space group C2/c matches with the standard JCPDS card No.80-1917. The intense and sharp diffraction spectrum indicates that well crystallized CuO nanostructures films can be obtained. This suggested that Fe atoms were effectively substituted by Cu sites within CuO lattice without affecting the crystal structure of copper oxide. No new phases were observed when increasing the Iron doping amount.

The variation of the average grain sizes (D) with doping ratio of the films were calculated from the peak widths, considering the most intense peak (-111) and using the Debye-Scherrer equation [25], is estimated from XRD patterns:

$$D = \frac{0.9\,\lambda}{\Delta\,\cos(\theta)} \quad (1)$$

Where: λ is the X-Ray wavelength (here the used value is 0.154 nm), Δ stands for the broadening of the diffraction line measured at half of its maximum intensity in radians and θ denotes the diffraction angle measured in degrees.

The results of this calculation are listed in Table 1. In fact, the average crystallite size of the pure CuO is found to be 88.27 nm and decreases slightly to 40.54 nm with increasing Fe doping ratio to 15%. The decrease of the grain size leads to the increase of nucleation centers density in the doping films giving rise to the formation of small crystallites.

The lattice parameters (a ≠ b ≠ c, α=ɤ=90°≠β) of the CuO monoclinic structure were calculated using the following equation [25-26]:

$$\frac{1}{d^2} = \frac{1}{\sin^2 \beta} \left( \frac{h^2}{a^2} + \frac{k^2 \sin^2 \beta}{b^2} + \frac{l^2}{c^2} - \frac{2hl \cos \beta}{ac} \right) \quad (2)$$

Where d is the interplanar distance and (hkl) are the Miller indices of the reflection plane.

The measured values of a, b, c are summarized in Table 1. These values are in good agreement with the reported values in the literature [27-30] and the standard JCPDS data card [31]. From table 1, it is observed that the variation of the lattice parameters are not remarkably affected comparing with the undoped CuO (x=0.00). This indicates that the doping with Fe ions has little influence on the crystalline structure. Hence, the combined action of Fe substitution and cation vacancy is responsible of such results [32-33].

In order to investigate the influence of Fe doping effect on the structural properties of our samples, the strain (ε) [34-35] and the dislocation density (δ) [36-37] have been calculated using the following relations:

$$\varepsilon = \frac{\Delta \cos \theta}{4} \quad (3)$$

The parameters Δ and θ are defined by the equation (1), so that:

$$\delta = \frac{1}{D^2} \quad (4)$$

The parameter D is also calculated using equation (1).

The calculated values are illustrated in Table 1. It is found that the values of strain are not affected when increasing the Fe concentrations. On the other hand, the dislocation density δ increases almost linearly as the concentrations of Fe-doped CuO increase as shown in Figure 2. Indeed, when increasing the Fe concentration the dislocation density increase inversely with the average grain size D due to small difference between Cu and Fe ion radii.

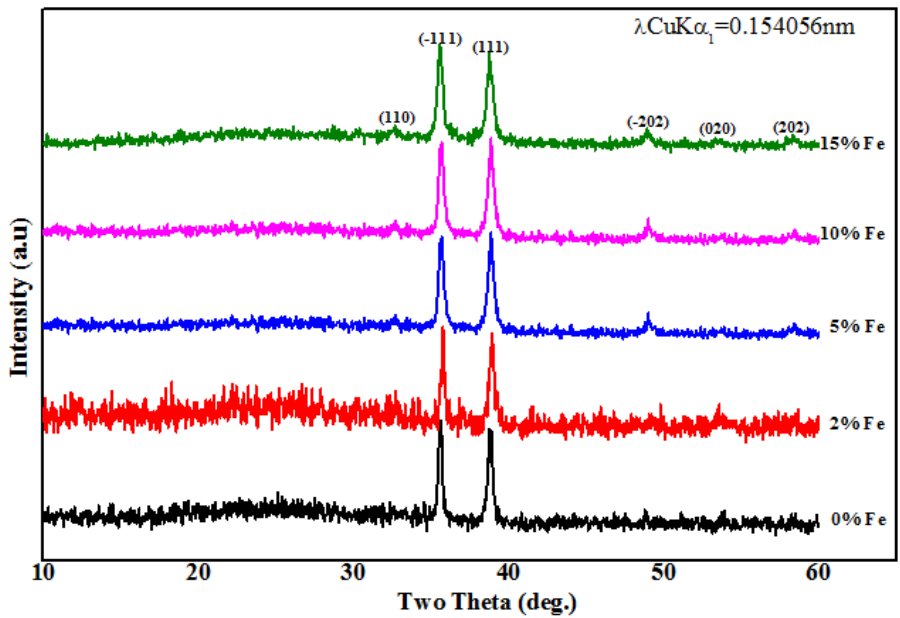

Figure 1. XRD diffractograms of Fe-doped CuO sprayed thin films

TABLE I. Structural parameters of Fe-doped CuO thin films

| $Cu_{1-x}Fe_xO$ Samples | Thickness (nm) | $2\theta$ (°) | Average crystallite size (D) (nm) | Lattice parameters | | | $\beta$ (°) | $\varepsilon$ (%) | $\delta$ ($10^{14}$ m$^{-2}$) |
| --- | --- | --- | --- | --- | --- | --- | --- | --- | --- |
| | | | | a (nm) | b (nm) | c (nm) | | | |
| x=0.00 | 681.28 | 35.60 | 88.27 | 0.479 | 0.336 | 0.523 | 99.46 | 0.2367 | 1.28 |
| x=0.02 | 689.25 | 35.76 | 65.38 | 0.478 | 0.335 | 0.522 | 99.47 | 0.2366 | 2.34 |
| x=0.05 | 670.93 | 35.63 | 54.35 | 0.477 | 0.344 | 0.520 | 99.48 | 0.2367 | 3.38 |
| x=0.10 | 546.20 | 35.65 | 45.88 | 0.469 | 0.343 | 0.514 | 99.50 | 0.2368 | 4.75 |
| x=0.15 | 448.80 | 35.56 | 40.54 | 0.468 | 0.342 | 0.513 | 99.52 | 0.2369 | 6.08 |

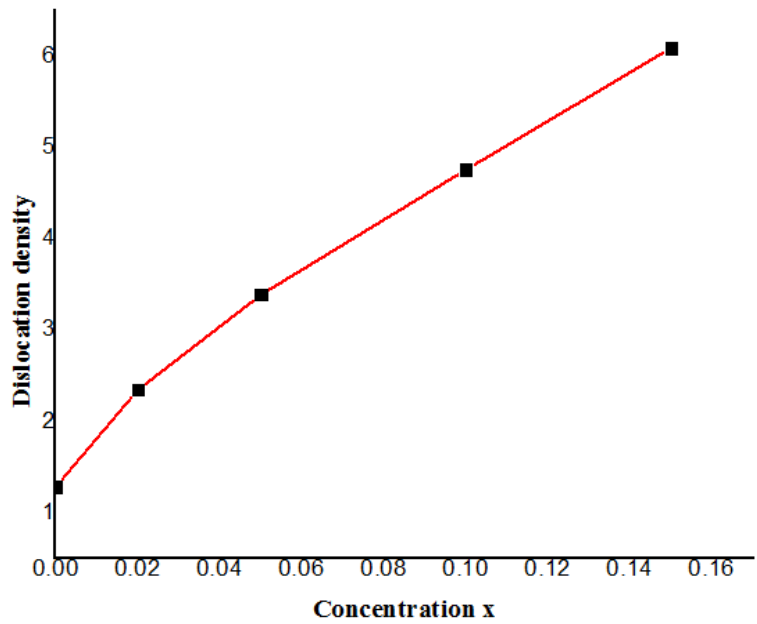

Figure 2. Dislocation density $\delta$ as a function of the Fe-doped CuO concentration x

### 3.2. Raman studies

In order to investigate the micro-structural properties and structural defects of our samples, the Raman spectroscopy was used. The Figure 3 depicts the FTIR Raman spectra of the prepared Fe doped CuO illustrating a number of vibrating modes in wave number ranging between 0 to 4000 cm$^{-1}$. The vibrational modes between 300 and 500 cm$^{-1}$ are related to the Cu-O of monoclinic CuO [38]. The absence of absorption modes related to secondary phases of the samples or other impurities confirms the successful doping with Fe ions in CuO [39, 40]. The Figure 4 illustrates the Raman spectra of the synthesized films. When doping with Iron atoms does not affect the monoclinic structure of pure CuO. In fact, the Raman spectra did not change when the concentration x increases from 0.00 to 0.15 of Fe doping, see Figure 4. This result can be attributed to particle size and defects in CuO crystal structure. Moreover, all peaks in Fe doped CuO nanostructures spectra are well matched with that of undoped nanostructures which corroborate well with XRD and FTIR results [41-42].

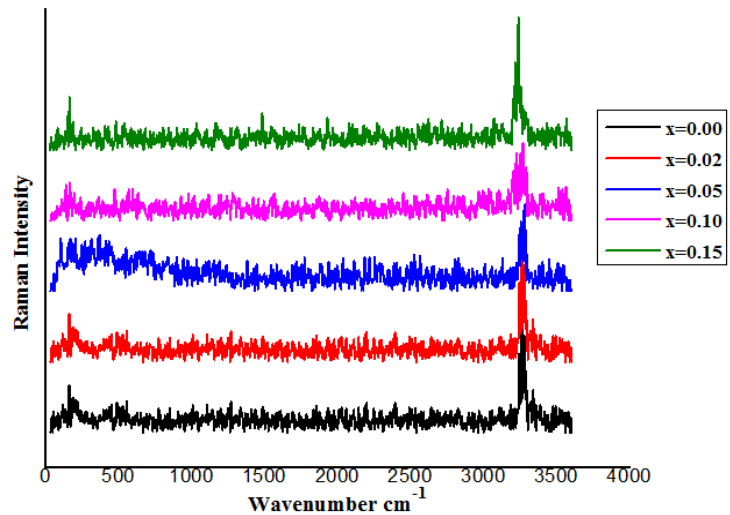

Figure 3. FTIR spectra of Fe-doped CuO thin films

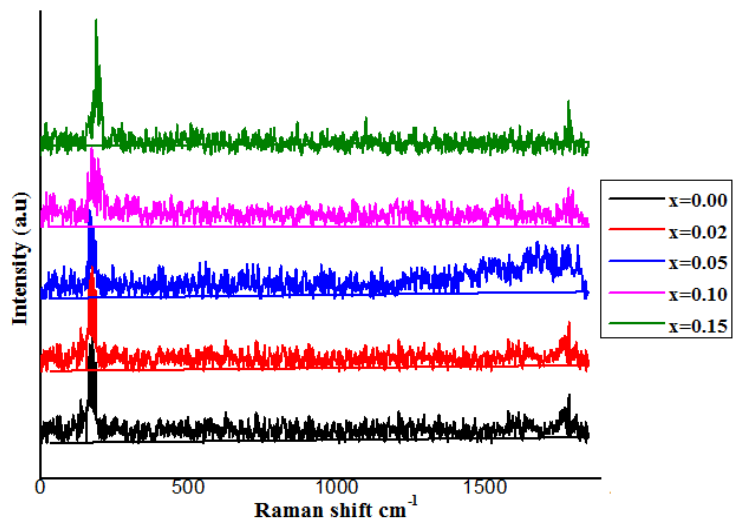

Figure 4. Raman spectra of Fe-doped CuO thin films

### 3.3. Optical properties

The optical properties the studied sprayed samples, were measured at room temperature using the UV/VIS/NIR spectrophotometer. The transmission was recorded in the range of 400 to 1400 nm for the incident beam, the obtained results are shown in Figure 5. As seen from this figure, the transmittance ($T_r$) of the films increases from 68% for pure CuO until 81% for the higher doping value of Fe (15%). On the other hand, it is found that for small wavelength values the films are dark compounds, whereas for higher values of the wavelength the films become transparent, see Figure 5. In addition, for a given wavelength, one can increase the transmittance by increasing the concentration of Iron. This result is similar to the results published by Oral et al. [43]. The thickness (t) values were illustrated to the Table 1, by using the Filmetrics F20, which decreases slightly.

The absorption coefficient ($\alpha_a$) [44] was calculated from Lambert's formula:

$$\alpha_a = \frac{1}{t}\ln\left(\frac{1}{T_r}\right) \qquad (5)$$

The optical band gap energy $E_g$ was estimated by extrapolating the linear part of the plots of $(\alpha_a h\upsilon)^2$ against the photon energy ($h\upsilon$) to the absorption coefficient ($\alpha_a=0$), as shown in Figure 6, using the following equation:

$$(\alpha_a h\upsilon)^2 = h\upsilon - E_g \qquad (6)$$

Where $\alpha_a$ is the absorption coefficient, $h\upsilon$ is the photon energy and $E_g$ is the bandgap energy.

The obtained value of the band gap energy of 1.29eV was for pure CuO. However, it was found to be 1.55, 1.86, 2.18 and 2.31 eV for $Cu_{1-x}Fe_xO$ (x=0.02, x=0.05, x=0.10, x=0.15), respectively. Similar results were reported by Sahay et al. for CuO nano-fibers [45], Basith et al. [46] and Mahmood et al. [47] revealed that the increase the band gap of the films has been effectively increases with increasing Fe contents.

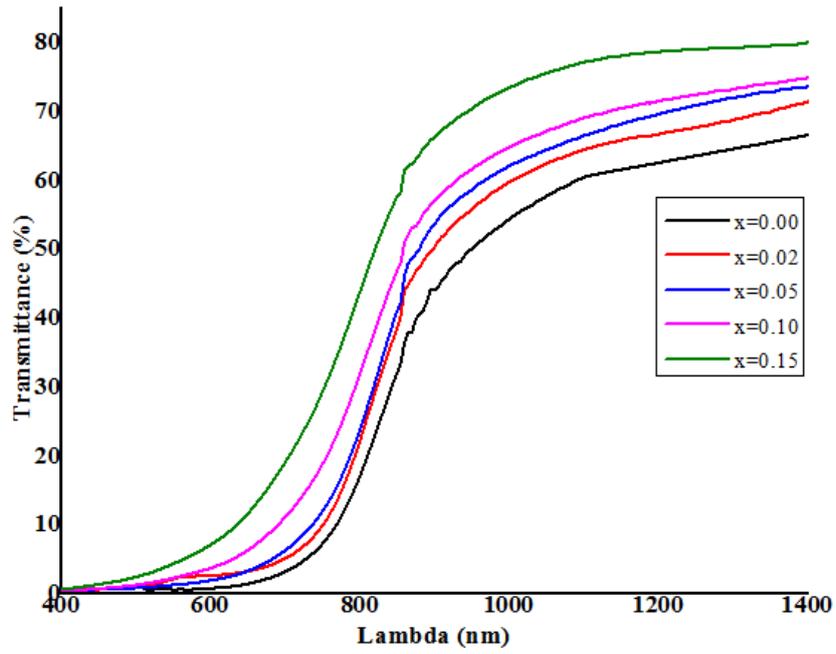

Figure 5.  The transmittance spectra of $Cu_{1-x}Fe_xO$ thin films

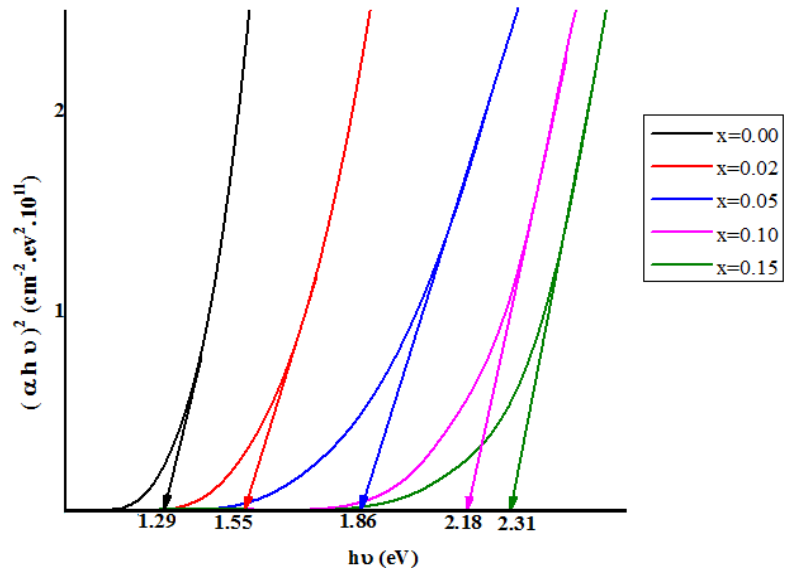

Figure 6.  The optical Band gap of $Cu_{1-x}Fe_xO$ thin films

## 3.4. Electrical properties

Using the Hall Effect measurements in the Van Der Pauw configuration at room temperature in air, the electrical properties of the formed particles were examined and their results are summarized in Table II. The results reveal that the resistivity of the films, prepared for the present study, decreases with increasing values of Fe doping content. This result is confirmed by Figure 7, which clearly indicates that the films have electrical properties. The XRD results confirm the same trend, which indicate that the films have a minimum resistivity at a higher concentration. This is probably due to the enhancement in the crystallinity of our films. Hence, an increasing of the electrical conductivity is found as it has been expected. Therefore, the conductivity can be expressed from the slope of the straight lines, as shown in the Arrhenius relationship [48-49]. The reason of such increase in the electrical conductivity can be explained by the existence of different active redox couples such as $Cu^{2+}/Cu^{+}$ (in pure CuO) together with $Fe^{2+}/Fe^{3+}$ which facilitates the electron hopping between $Cu^{2+}/Fe^{2+}$ and/or $Cu^{+}/Fe^{3+}$ ion pairs as a result of mutual charge interaction [50-55]. On the other hand, the increase in the electrical conductivity may correspond to a possible facility in the mobility of charge carriers. Indeed, the growth of the carrier mobility is due to Fe doping. In fact, the oxygen insertion can enhance the degree of the lattice distortion and thus strengthen the carrier scattering [56-57]. Consequently, the majority carrier concentration is obtained in the order of $10^{17}$ cm$^{-3}$ and decreases as shown in Table 2. Such decrease of carrier concentration facilitates the carrier mobility, giving rise to an enhancement of p-type conductivity by more than one order of magnitude [58]. All the samples showed p-type conductivity.

TABLE II. Electrical parameters of Fe-doped CuO thin films

| $Cu_{1-x}Fe_xO$ Samples | Resistivity $\rho$ ($\Omega$cm) | Conductivity $\sigma$ ($\Omega^{-1}$cm$^{-1}$) | Mobility $\mu$ (x10 cm$^2$V$^{-1}$s$^{-1}$) | Carrier concentration n (x10$^{17}$cm$^{-3}$) | Carrier type (Hall Effect) |
|---|---|---|---|---|---|
| x=0.00 | 7.530 | 0.1328 | 11.9 | 0.976 | P |
| x=0.02 | 5.060 | 0.1976 | 15.3 | 0.642 | P |
| x=0.05 | 3.290 | 0.3039 | 15.9 | 0.619 | P |
| x=0.10 | 0.315 | 3.1746 | 21.3 | 0.422 | P |
| x=0.15 | 0.063 | 15.873 | 60.5 | 0.192 | P |

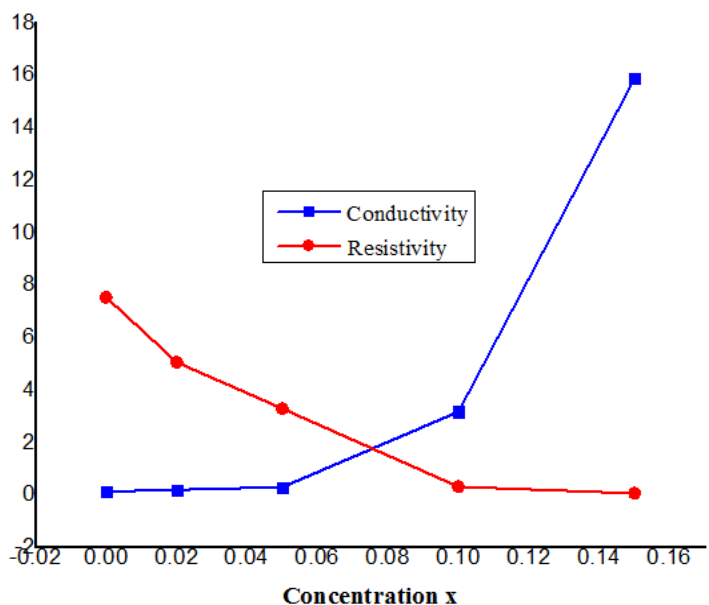

Figure 7. The electrical parameters of $Cu_{1-x}Fe_xO$ thin films

## 4. Conclusion

The structural, micro-structural, optical and electrical structures for the $Cu_{1-x}Fe_xO$ compound were investigated using the XRD, FTIR-Raman spectroscopy, UV-VIS spectrometer and Hall Effect measurements, respectively. In one hand, the X-Ray diffraction (XRD) pattern showed that Fe doping contents improve the crystallinity of CuO without deteriorating its monoclinic structure. Raman spectroscopy results have confirmed that the phase purity is not affected for the synthesized films. On the other hand, the films with highest Fe doping concentration, showed highest transmittance percentage values reaching 80 %. The increase of the optical band gap when increasing the Fe doping amount might be explained by the decrease of the grain sizes. Hence, the electrical part unveiled a significant increase in the conductivity and the mobility within an increase in Fe-doping content. A decrease in the resistivity and the carrier concentration of the films is investigated. It confirmed that the growth in the films is p-type in nature.